\documentclass[aps,preprint, floatfix,nofootinbib,showpacs]{revtex4}
\usepackage{graphicx,epstopdf,color}
\usepackage{graphicx,epstopdf}
\begin{document}

\title{Longitudinal Weak Gauge Bosons Scattering in Hidden $Z'$ Models}
\renewcommand{\thefootnote}{\fnsymbol{footnote}}

\author{
Kingman Cheung$^{1,2,3}$, Cheng-Wei Chiang$^{4,5}$, Yu-Kuo Hsiao$^5$,
and Tzu-Chiang Yuan$^5$}
\affiliation{
$^1$Division of Quantum Phases \& Devices, School of Physics,
Konkuk university, Seoul 143-701, Korea\\
$^2$ Physics Division, National Center for Theoretical Sciences, Hsinchu, Taiwan\\
$^3$ Department of Physics, National Tsing Hua University,
Hsinchu, Taiwan \\
$^4$ Department of Physics and Center for Mathematics and Theoretical
Physics, National Central University, Chung-Li, Taiwan \\
$^5$ Institute of Physics, Academia Sinica, Nankang, Taipei 11925, Taiwan
}
\renewcommand{\thefootnote}{\arabic{footnote}}
\date{\today}

\begin{abstract}
  Longitudinal weak gauge boson scattering has been well known as a powerful method to probe the underlying mechanism of the electroweak symmetry breaking sector of the Standard Model.  We point out that longitudinal weak gauge boson scattering is also sensitive to the gauge sector when the non-Abelian trilinear and quartic couplings of the Standard Model $Z$ boson are modified due to the general mixings with another $Z'$ boson
in the hidden sector and possibly with the photon as well.  In particular, these mixings can lead to a partially strong scattering effect in the channels of $W^\pm_L W^\pm_L \to W^\pm_L W^\pm_L$ and $W^\pm_L W^\mp_L \to W^\pm_L W^\mp_L$ which can be probed at the Large Hadron Collider.  We study this effect in a simple $U(1)$ extension of the Standard Model recently suggested in the literature that includes both the symmetry breaking Higgs mechanism as well as the gauge invariant Stueckelberg mass terms for the two Abelian groups.  Other types of $Z'$ models are also briefly discussed.
\end{abstract}
\maketitle


\section{Introduction}

The CERN Large Hadron Collider (LHC) will soon be reactivated after the year 2008 accident
to uncover the mystery of electroweak symmetry breaking (EWSB).  The ultimate goal of the LHC is to search for the Higgs boson and hopefully any new physics beyond the Standard Model (SM).  Many grand-unified theories, extra-dimensional models as well as string-inspired models predict additional $U(1)$ gauge groups in addition to the SM hypercharge $U(1)_Y$.  Therefore, at least one extra heavy neutral gauge boson $Z^\prime$ is generally expected in these theories.  There have been many studies of $Z'$ bosons at colliders \cite{zprime}.  The most direct channel to probe the existence of $Z'$ boson is the Drell-Yan process, which will be identified unambiguously by a new resonance peak in the invariant mass distribution of the electron-positron or muon-antimuon pairs.  The current best limit of this search is from the Tevatron \cite{tevatron-limit}.  The lower mass limits of a few popular $Z'$ models range from $0.7-1.0$ TeV.  One can also measure the branching ratios of the $Z'$ to differentiate the underlying models, as was studied recently in Ref.~\cite{godfrey}.  At the LHC, it has been shown that one can probe a $Z'$ boson up to about a few TeV.  Thus, if the $Z'$ boson is above a few TeV or has suppressed couplings to electrons and muons, the LHC may not be able to identify its presence easily.

In this work, we use the longitudinal vector boson scattering \cite{WW,equiv,madison} to probe the gauge sector.  We show that if the SM $Z$ boson mixes with a heavy enough $Z'$ boson of any origin such that the gauge coupling of the SM $Z$ boson to a pair of $WW$ is modified, the longitudinal vector boson scattering will show an appreciable rise in scattering cross sections.  This can be understood as follows.  Consider the SM first for the channel $W_L^+ W_L^- \to W_L^+ W_L^-$.  Besides the contributions from the $\gamma,Z$, and the Higgs exchange diagrams, there is also the 4-point contact interaction diagram.  Recall that at the asymptotically high energy limit, the longitudinal polarization $\epsilon^\mu(p)$ for the $W$ boson behaves like $p^\mu/M_W$.  Therefore, naively the 4-point diagram goes like $(E/M_W)^4$ as $E \gg M_W$ where $E$ is the center-of-mass (CM) energy.  Such bad energy-growing terms will be offset, however, by similar terms from the pure gauge diagrams with $\gamma$ and $Z$ exchanges, leaving behind those $(E/M_W)^2$ terms, which will eventually be cancelled by the Higgs diagrams.  Only the $(E/M_W)^0$ terms survive and unitarity is guaranteed in the SM.  In a recent work by three of the authors \cite{us}, we show that in extended Higgs models if the light Higgs boson has a reduced coupling $g_{hWW}$ and the heavy Higgs boson is heavy enough, there will be a wide energy range in which the longitudinal vector boson scattering becomes strong and detectable at the LHC.  The general two-Higgs-doublet model is an example of such a scenario.  The energy-growing terms of order $(E/M_W)^2$ exist and cause the rising of the scattering cross sections between the mass scales of light and heavy Higgs bosons.  In the present work, we show even more spectacular rising in the scattering cross section due to a modified $g_{WWZ}$ coupling such that the energy-growing terms are effectively of order $(E/M_W)^4$.  The modified coupling $g_{WWZ}$ arises from the mixing between the SM $Z$ boson and an extra $Z'$ boson of some origin.  The mixing can be of the kinetic type \cite{Holdom, Goldberg-Hall} or the Stueckelberg type \cite{Kors-Nath,Feldman-Liu-Nath-1,Feldman-Liu-Nath-2}
or the combination of both \cite{Feldman-Liu-Nath-3}.  We shall consider mainly the Stueckelberg model and comment briefly on the kinetic mixing and a few other types of $Z'$ models.

The organization of this paper is as follows.  In the next section,
we describe the Stueckelberg $Z'$ model and how the mixing could lead to the modified trilinear and quartic gauge couplings as well as the gauge-Higgs couplings which are relevant to the $W_L W_L$ scatterings.
We also work out a mass relation between $Z$, $Z'$ and $W$ arising from the custodial symmetry of the model. This mass relation is crucial for the restoration of unitarity at very high energy in the $W_LW_L$ scattering amplitudes.
In Sec.~III,  we first remind the readers of some details about various contributions to the scattering amplitude of $W^+_L W^-_L \to W^+_L W^-_L$ in the SM. The exact scattering amplitude for $W^+_L W^-_L \to W^+_L W^-_L$ with modified couplings in the Stueckelberg $Z'$ model as well as
its high energy limits will be presented.  We present our numerical results in Sec.~IV and conclude in Sec.~V.


\section{The Stueckelberg Extension of Standard Model}

The Stueckelberg extension of the SM (StSM) \cite{Kors-Nath} is obtained by adding a hidden sector associated with an extra $U(1)_C$ interaction, under which the SM particles are neutral.  Assuming there is no kinetic mixing between the two $U(1)$'s, the Lagrangian describing the system is ${\cal L} ={\cal L}_{\rm SM} + {\cal L}_{\rm StSM}$, with
\begin{eqnarray}
{\cal L}_{\rm SM} &=& -\; \frac{1}{4} W_{\mu\nu}^a \, W^{a\mu\nu}
                      - \frac{1}{4} B_{\mu\nu}\, B^{\mu\nu} + i \bar f \gamma^\mu D_\mu f
                   + D_\mu \Phi^\dagger \, D^\mu \Phi - V(\Phi^\dagger \,\Phi) \; , \\
{\cal L}_{\rm StSM} &=&
         - \frac{1}{4} C_{\mu\nu}\, C^{\mu\nu}
 + \frac{1}{2} \left( \partial_\mu \sigma + M_1 C_\mu + M_2 B_\mu \right )^2
  +  {\overline \chi} \left( i \gamma^\mu D^X_\mu - M_\chi \right) \chi ~,
\label{stusm}
\end{eqnarray}
where $W_{\mu\nu}^a(a=1,2,3)$, $B_{\mu\nu}$, and $C_{\mu\nu}$ are the field strength tensors of the gauge fields $W_\mu^a$, $B_\mu$, and $C_\mu$, respectively; $f$ denotes a SM fermion, while $\chi$ is a Dirac fermion in the hidden sector which may play a role as milli-charged dark matter in the
Universe \cite{Feldman-Liu-Nath-3,Cheung-Yuan} and $M_\chi$ is its mass; $\Phi$ is the SM Higgs doublet;
and $\sigma$ is the Stueckelberg axion scalar.  The covariant derivatives $D_\mu = (\partial_\mu + i g_2 \vec T \cdot {\vec W}_\mu + i g_Y\frac{Y}{2} B_\mu )$ and $D^X_\mu = (\partial_\mu + i g_X Q^\chi_X C_\mu )$.

The mass term for $V \equiv ( C_\mu,\, B_\mu,\, W^3_\mu )^T$, after the EWSB $\langle \Phi \rangle = v/\sqrt{2}$ with a vacuum expectation value $v \simeq 246$ GeV, is given by
\begin{equation}
- \frac{1}{2} V^{\mathrm T}  M_{\mathrm{Stu}}^2  V  \equiv   -\frac{1}{2}
  \left ( C_\mu,\, B_\mu,\, W^3_\mu \right ) \;
 \left( \begin{array}{ccc}
  M_1^2 & M_1 M_2 & 0 \\
 M_1 M_2 & M_2^2  + \frac{1}{4} g_Y^2 v^2 & -\frac{1}{4} g_2 g_Y v^2 \\
 0 & -\frac{1}{4} g_2 g_Y v^2  & \frac{1}{4} g_2^2 v^2 \end{array} \right ) \;
 \left( \begin{array}{c}
       C_\mu \\
       B_\mu \\
       W^3_\mu \end{array} \right ) \;.
\end{equation}
One can easily show that the determinant of $M_{\mathrm{Stu}}^2$ is zero, indicating the existence of at least one zero eigenvalue to be identified as the photon mass.  A similarity transformation can bring the mass matrix $M^2_{\mathrm{Stu}}$ into a diagonal form
\begin{equation}
\left( \begin{array}{c}
                                        C_\mu \\
                                        B_\mu \\
                                        W^3_\mu \end{array} \right ) =
  O \, \left( \begin{array}{c}
                                        Z'_\mu \\
                                        Z_\mu \\
                                        A_\mu \end{array} \right )
\;\; , \;\;\;\;\;
 O^{\mathrm T}  M_{\mathrm{Stu}}^2 \, O \, =
 \mathrm{Diag} \left[ M^2_{Z'},\, M^2_Z,\, 0 \right]  \; .
 \label{rotate}
\end{equation}
Hereinafter, we denote $A$, $Z$ and $Z'$ as the physical mass eigenstates.  Define
\begin{equation}
M_Y = \frac{1}{2} g_Y v \; .
\label{ymass}
\end{equation}
Then the $Z$ and $Z'$ masses can be written as
\begin{eqnarray}
M^2_{Z',\,Z} &=& \frac{1}{2} \biggr[
 M_1^2 + M_2^2 + M_W^2 + M_Y^2 \nonumber\\
&& \pm \sqrt{ (M_1^2 + M_2^2 +  M_W^2 + M_Y^2 )^2
 - 4 ( M_1^2 + M^2_2) M_W^2 - 4 M_1^2 M_Y^2   }  \biggr ]\;,
\label{zmass}
\end{eqnarray}
where we have used $M_W = g_2 v/2$.  The orthogonal matrix $O$ is parameterized as
\begin{equation}
O_{ij} = \left( \begin{array}{ccc}
              c_\psi c_\phi - s_\theta s_\phi s_\psi \;\; &
              s_\psi c_\phi + s_\theta s_\phi c_\psi \;\; &
            - c_\theta s_\phi \\
              c_\psi s_\phi + s_\theta c_\phi s_\psi \;\;&
              s_\psi s_\phi - s_\theta c_\phi c_\psi \;\;&
              c_\theta c_\phi \\
            - c_\theta s_\psi & c_\theta c_\psi & s_\theta \end{array}\right )
 \;,
\label{matrixOrth1}
\end{equation}
where $s_\phi =\sin\phi, c_\phi=\cos\phi$ and similarly for the angles $\psi$ and $\theta$.  The angles are related to the original parameters in the Lagrangian ${\cal L}$ by
\begin{equation}
\epsilon \equiv \tan \phi = \frac{M_2}{M_1} \; ,
\label{mixing-angle-1}
\end{equation}
\begin{equation}
 \tan\theta = \frac{g_Y }{g_2} \cos\phi \; , \;\;\; e = g_2 \sin \theta \; ,
\label{mixing-angle-2}
\end{equation}
\begin{equation}
 \tan\psi = \frac{ M_W^2 \tan\theta\, \tan\phi }
                 {\cos\theta [ M^2_{Z'} - M_W^2 (1+ \tan^2\theta) ]} \; .
\label{mixing-angle-3}
\end{equation}


\subsection{Modified Couplings}

In this model, the couplings between the neutral gauge bosons and
the Higgs are given by
\begin{eqnarray}
\label{HVVCouplings}
{\cal L}_{{\rm Higgs}-Z-Z^\prime} & = &
\frac{1}{8} \left(H^2 + 2 v H \right)
\left[
\left( g_2 O_{32} - g_Y O_{22}  \right)^2 Z_\mu Z^\mu +
 \left( g_2 O_{31} - g_Y O_{21} \right)^2 Z^\prime_\mu Z^{\prime\mu} \right. \nonumber \\
&+ & \left. 2 \left( g_2 O_{31} - g_Y O_{21}  \right) \left( g_2 O_{32} - g_Y O_{22}  \right)  Z_\mu Z^{\prime\mu} \right] \; .
\end{eqnarray}
Feynman rules for the $HVV$ and $HHVV$ couplings can be read off easily from Eq.~(\ref{HVVCouplings}).  We note that due to Eq.~(\ref{mixing-angle-2}), there are no $H\gamma\gamma$, $H\gamma Z$ and $H\gamma Z^\prime$ couplings from the mixings as one should expect by the fact that photon must couple to the fields with nonzero electric charges at tree level.

The modified trilinear and quartic pure gauge couplings in this model can be derived in a straightforward way, and they are given by
\begin{equation}
W(k_1,\mu)W(k_2,\nu)V_i(k_3,\lambda) \; : \; - i g_2 O_{3i}
[(k_1-k_2)_\lambda g_{\mu\nu}+(k_2-k_3)_\mu g_{\nu\lambda}+(k_3-k_1)_\nu g_{\lambda\mu}]
\end{equation}
\begin{equation}
W(k_1,\mu)W(k_2,\nu)V_i(k_3,\lambda)V_j(k_4,\rho) \; : \; - i g_2^2 O_{3i} O_{3j}
[2g_{\mu\nu}g_{\lambda\rho}-g_{\mu\lambda}g_{\nu\rho}-g_{\mu\rho}g_{\nu\lambda}]
\end{equation}
where $V_i = Z', Z, A$ for $i = 1,2,3$ respectively.
%


\subsection{Mass Relation}

In SM, we have
\footnote{This equation is also true in StSM.}
\begin{equation}
\label{wmassdef}
M_W = \frac{1}{2}g_2 v \; .
\end{equation}
The custodial SU(2) symmetry protects the following tree level symmetry breaking mass relation
\begin{equation}
\label{mr1}
M_W = M_Z \cos \theta_W \;
\end{equation}
from receiving large radiative corrections.
The above mass relation is essential for the cancellation of the bad $E^4$ terms
in the $WW$ scattering amplitude in order to maintain unitarity at high energy.
In StSM, a similar mass relation among $M_W$, $M_Z$ and $M_{Z'}$ exists in order
to tame the bad high energy terms and it is given by
\footnote{We note that this mass relation holds even when kinetic
mixing is allowed between the two abelian gauge groups.}
\begin{eqnarray}
\label{mr2}
M_W^2 & = & \cos^2 \theta \left(  M_Z^2 \cos^2 \psi+  M_{Z'}^2 \sin^2 \psi \right) \; .
\end{eqnarray}
In the custodial symmetry limit of $g_Y \to 0$, it is easy to show that $M_Z = M_W$ and
$M_{Z'} = \left(M_1^2 + M_2^2 \right)^{1/2}$. Hence in this model, the $W^+, W^-, Z$ form a triplet of the custodial $SU_L(2) \times SU_R(2)$ just like the case in SM, while $Z'$ transforms as a singlet. The mass relation Eq. (\ref{mr2}) is trivially satisfied
by setting $\theta$ (and hence $\psi$ from Eq. (\ref{mixing-angle-3})) to be zero.

\begin{figure}[th!]
\centering
\includegraphics[width=3in]{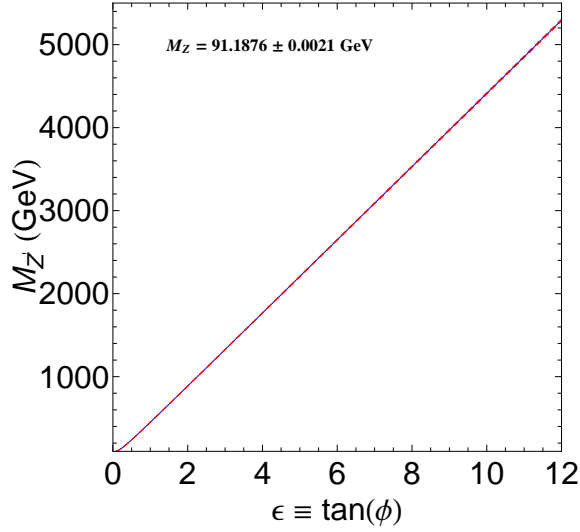}
\caption{Contour of the $Z$ mass as a function of $\tan ( \phi )$ and $M_{Z'}$.
The solid blue line corresponds to the central value of measured $M_Z$ and the red dashed lines correspond to its
$1\sigma$ deviation.}
\label{zmasscontour}
\end{figure}

Another useful formula for the $Z$ mass is
\begin{eqnarray}
\label{zmass4contour}
M_Z^2 & = & \frac{ M_W^2 \left( M_{Z'}^2 - M_W^2 - M_Y^2 \right)} { M_{Z'}^2 \cos^2\theta - M_W^2} \;\; , \\
& =& \frac{ M_W^2 \left( M_{Z'}^2 - M_W^2
- \frac{1}{4}  \frac{e^2} { \cos^2 \theta}  \left( 1 + \epsilon^2 \right) v^2  \right) } {M_{Z'}^2 \cos^2 \theta - M_W^2} \;\; .
\end{eqnarray}
In Fig. \ref{zmasscontour}, we plot the contour for the $Z$ mass as a function of $\epsilon$ and $M_{Z'}$
using the above formula.
Input parameters for this plot are $M_W$, $\alpha_{em}(M_Z)$ and the Fermi constant $G_F$.
Other quantities are fixed by
$v = \left( {\sqrt 2} G_F \right)^{-1/2}$, $g_2 = 2 M_W / v$ and $\sin \theta = e /g_2$.
It is clear that the experimental value of $Z$ mass by itself does not exclude the possibility of large $\epsilon$ as long as an appropriate large ${Z'}$ mass is also chosen. Such a large angle scenario is necessary
for the energy growing terms of order $(E/M_W)^4$ in the $W_L W_L$ scatterings to give rise to interesting physical effects since these terms are hampered by the mixing angles in their coefficients. We will show this more explicitly in the next section.

\subsection{Neutral Current Interactions}

The interactions of fermions with the neutral gauge bosons before rotating
to the mass eigenbasis are given by
\begin{equation}
- {\cal L}^{NC}_{\rm int} =
  g_2 W^3_\mu \, \bar f \gamma^\mu T^f_3 f
+ g_Y B_\mu \, \bar f \gamma^\mu \frac{Y^f}{2} f
+ g_X C_\mu \, {\overline \chi} \gamma^\mu Q^\chi_X \chi  \;,
\end{equation}
where $f$ denotes the SM fermions.  The neutral gauge fields are rotated into the mass eigenbasis using Eq.~(\ref{rotate}), and the above neutral current interaction becomes
\begin{eqnarray}
- {\cal L}^{NC}_{\rm int} &=& \bar f\, \gamma^\mu \left[
  \left( \epsilon_{Z'}^{f_{L}} P_L + \epsilon_{Z'}^{f_{R}} P_R\right)\, Z'_\mu
+ \left( \epsilon_{Z}^{f_{L}} P_L + \epsilon_{Z}^{f_{R}}P_R \right)\, Z_\mu
+ e Q_{\rm em} A_\mu \right ] f \nonumber\\
&+&
 {\overline \chi} \gamma^\mu \left[
   \epsilon^\chi_\gamma A_\mu
 +\epsilon^\chi_Z Z_\mu
 +\epsilon^\chi_{Z'} Z'_\mu \right ]\, \chi \; ,
\label{chiral1}
\end{eqnarray}
where $\epsilon^\chi_{V_i} = \tilde g^\chi_X O_{1i}$ with $\tilde g^\chi_X \equiv g_X Q^\chi_{X}$ and
\begin{eqnarray}
 \epsilon_Z^{f_L} &=& \frac{g_2}{\cos\theta} \cos\psi
 \left[ \left( 1 -  \epsilon \sin \theta \tan \psi \right) T^3_f
 - \sin^2\theta \left( 1 -  \epsilon \csc \theta \tan \psi \right) Q_f \right] \; , \nonumber\\
 \epsilon_Z^{f_R} &=& - \frac{g_2}{\cos\theta} \cos\psi
 \sin^2\theta \left( 1 -  \epsilon \csc \theta \tan \psi \right) Q_f \; , \nonumber\\
 \epsilon_{Z'}^{f_L} &=& - \frac{g_2}{\cos\theta} \cos\psi
 \left[ \left( \tan \psi +  \epsilon \sin \theta \right) T^3_f
 - \sin^2\theta \left(  \epsilon \csc \theta + \tan \psi \right) Q_f \right] \; , \nonumber\\
 \epsilon_{Z'}^{f_R} &=& \frac{g_2}{\cos\theta} \cos\psi
  \sin^2\theta \left(  \epsilon \csc \theta + \tan \psi \right) Q_f \;.
 \label{chiral2}
\end{eqnarray}

In the SM limit, these couplings reduce to the usual formulas
\begin{eqnarray}
 \epsilon_Z^{f_L}(SM) &=& \frac{g_2}{\cos\theta_W}  \left[ T^3_f - \sin^2\theta_W Q_f \right] \; , \nonumber\\
 \epsilon_Z^{f_R}(SM) &=& - \frac{g_2}{\cos\theta_W} \sin^2\theta_W  Q_f \; ,\nonumber\\
 \epsilon_{Z'}^{f_{L(R)}}(SM)  &=& 0 \; ,
 \label{chiralSM}
\end{eqnarray}
in terms of the Weinberg angle $\theta_W$.
Since the couplings of the StSM in Eq.~(\ref{chiral2}) could shift from those of the SM in Eq.~(\ref{chiralSM}),
we will examine in the next subsection their validity in the large $\tan \phi$ scenario
by using the measured electroweak quantities in $Z$ decays.

\subsection{$Z$ decays in the StSM of large $\tan \phi$}

To be specific, let's first consider the decay of $Z\to e^+ e^-$ as an example.
As for the inputs, we take $M_{Z'}=1$ TeV and $\sin^2\theta=e^2/g_2^2=0.229$ in Eq.~(\ref{mixing-angle-2}) for the StSM,
while $\sin^2\theta_W=0.231$ for the SM is adopted from PDG \cite{pdg}.
We note that $\sin^2\theta= \sin^2\theta_W$ is only valid for small $\tan\phi$.
With these inputs and using Eqs.~(\ref{mixing-angle-3}) and (\ref{mr2}), we find
$\epsilon = 2$ and $\tan \psi = 0.008$.
Thus we obtain the ratio of $\epsilon_Z^{e_R}$ to $\epsilon_Z^{e_R}(SM)$,
\begin{eqnarray}
\frac{\epsilon_Z^{e_R}}{\epsilon_Z^{e_R}(SM)}&=&\frac{\sin^2\theta \cos\psi/\cos\theta}{\sin^2\theta_W /\cos\theta_W}\left( 1 -  \epsilon \csc \theta \tan \psi \right)\nonumber\\
&=&1.078\times (1-0.081)\nonumber\\
&=&0.991\,.
\end{eqnarray}
According to this simple calculation, the 8\% change by the term $\epsilon \csc \theta \tan \psi$
compared to 1 for the large $\tan\phi$ scenario considered in our work,
is alleviated by the prefactor,
such that $\epsilon_Z^{e_R}$ does not deviate too much from its SM value.
In the same way, the deviation of $\epsilon_Z^{e_L}$ from $\epsilon_Z^{e_L}(SM)$ is checked to be of $O(1\%)$.

We further check other quantities like $\Gamma_Z$, $\Gamma(\textrm{had})$, $\Gamma(\ell^+ \ell^-)$, $R_e$, $R_b$, $R_c$, $A_b$ and $A_c$
defined in the PDG, and they all present an $O(1\%)$ modification compared to their SM values
which are tolerable for the tree level calculation (See Table \ref{zdecays}).
We note that the quantity $A_e$ is very sensitive to the values of $\sin^2\theta_W$ in different schemes, and so we ignore $A_e$ in our tree level treatment.
\begin{table}[b]
\caption{Quantities in $Z$ decays. The experimental data are in the second column,
while the SM and the StSM of the tree-level calculation are in the third and fourth columns, respectively. \label{zdecays}}
\begin{ruledtabular}
\begin{tabular}{l|c|c|c}
Quantity & Experimental Data & SM & StSM\\
\hline
$\Gamma_Z$[GeV]       &$2.4952 \pm 0.0023 $    &2.4226  &2.4261\\
$\Gamma(\textrm{had})$[GeV]    &$1.7444 \pm 0.0020 $    &1.6747  &1.6824\\
$\Gamma(\ell^+ \ell^-)$[MeV]&$83.984 \pm 0.086  $    &83.415  &83.292\\
$\sigma_{\textrm{had}}$[nb]    &$41.541 \pm 0.037  $    &42.022  &42.031\\
$R_e$                 &$20.804 \pm 0.050  $    &20.077  &20.198\\
$R_b$                 &$0.21629\pm 0.00066$    &0.2197  &0.2193\\
$R_c$                 &$0.1721 \pm 0.0030 $    &0.1704  &0.1710\\
$A_b$                 &$0.923  \pm 0.020  $    &0.936   &0.941\\
$A_c$                 &$0.670  \pm 0.027  $    &0.669   &0.697\
\end{tabular}
\end{ruledtabular}
\end{table}


\section{scattering amplitudes}

\subsection{General Consideration}

As a concrete example, consider the scattering of $W^+_L W^-_L \to W^+_L W^-_L$, which proceeds through the $s$- and $t$-channels of $\gamma$ and $Z$ exchanges, the $s$- and $t$-channels of Higgs exchanges, and the 4-point seagull diagram.  At high energies the longitudinal polarization $\epsilon^\mu_L(p)$ of the $W$ boson is proportional to its momentum $p^\mu$, and it can be expressed as $\epsilon^\mu_L(p) = p^\mu / M_W + v^\mu(p)$ with a small correction $v^\mu(p) \simeq -\left[ M_W /(2 (p^0)^2) \right] ( p^0, - {\bf p}) \sim O(M_W/E_W)$.  In the CM system of $W^+_L (p_1) W^-_L (p_2) \to W^+_L(k_1) W^-_L(k_2)$, one can choose $v^\mu(p_1) = - 2 (M_W/s ) p_2^\mu$, and so on.  In the high energy limit, the amplitudes for the 4-point seagull diagram and the $\gamma,Z$ exchange diagrams in $s$- and $t$-channels are given by
\begin{eqnarray}
\label{m4}
i {\cal M}_4 &\simeq& i \frac{g_2^2}{4M_W^4} \left[ s^2 + 4 st + t^2 - 4 M_W^2(s+t)
 - \frac{8 M_W^2}{s} u t \right ] \; , \\
 \label{ms-sm}
i {\cal M}_s^{\gamma+Z} &\simeq& -i \frac{g_2^2}{4 M_W^4} \left[ (t-u)(s+4M_W^2)\right ]s\,A^{SM}_s \;, \\
\label{mt-sm}
i {\cal M}_t^{\gamma+Z} &\simeq& -i \frac{g_2^2}{4 M_W^4} \left[ (s-u)(t-4M_W^2)-\frac{8 M_W^2}{s} t^2 \right ]t\,A^{SM}_t \; ,
\end{eqnarray}
where $A^{SM}_{x}$ with $x = s$ or $t$ arises from the propagator factor and is given by
\begin{eqnarray}\label{AxSM}
A^{SM}_{x}&=&\frac{\text{sin}^2\theta_W}{x}+\frac{\text{cos}^2\theta_W}{x-M_Z^2}
=\frac{1}{x} \left(1+\frac{M_Z^2 \text{cos}^2\theta_W}{x-M_Z^2} \right)\,.
\end{eqnarray}
Note that each of these individual amplitudes contains terms proportional to $E^4/M_W^4$ where $E$ is the CM energy, as one would naively expect from the form of the longitudinal polarization of the $W$ boson.  However, the gauge structure of the SM guarantees the cancellation of the $E^4/M_W^4$ terms.  All one is left with are the $E^2/M_W^2$ terms in the high energy limit.  The sum of the amplitudes of the pure gauge diagrams
$i {\cal M}^{\rm gauge}=i({\cal M}_4+{\cal M}_s^{\gamma+Z}+{\cal M}_t^{\gamma+Z})$ is therefore
\begin{eqnarray}
\label{amp-gauge}
i {\cal M}^{\rm gauge} & \simeq & - i \frac{g_2^2}{4 M_W^4}\, \bigg[4M_W^2-3(M_Z^2\text{cos}^2\theta_W)\bigg]u + O\left((E/M_W)^0\right)\;
\nonumber \\
& \simeq & - i \frac{g_2^2}{4 M_W^2}\, u + O\left((E/M_W)^0\right) \;,
\end{eqnarray}
where the custodial $SU(2)$  mass relation $M_W^2 = M_Z^2 \cos^2 \theta_W$ has been used.
On the other hand, the sum of the two diagrams from Higgs exchange is
\begin{eqnarray}
\label{amp-Higgs}
i {\cal M}^{\rm Higgs} &\simeq& - i \frac{g_2^2}{4 M_W^2} \left[
  \frac{ (s - 2 M_W^2)^2}{s -M_h^2} + \frac{(t - 2 M_W^2)^2}{t -M_h^2}
  \right ] \;  \nonumber \\
 & \simeq &
  i \frac{g_2^2}{4 M_W^2} \, u  \;,
\end{eqnarray}
in the limit of $s \gg M_h^2, M_W^2$.  Thus, the bad energy-growing term is delicately cancelled between the gauge diagrams and the Higgs diagrams.  This is a well-known fact in the SM.

Suppose there exists a heavy $Z'$ boson that mixes with the SM $Z$ boson.  The one observed at LEP is the lighter mass eigenstate, which is mostly the SM one: $Z_1 = \cos\theta_{Z Z'} Z + \sin\theta_{Z Z'} Z'$, where $\theta_{ZZ'}$ is a small mixing angle.  Naively, the $g_{WWZ}$ coupling is modified by a multiplicative factor $\cos\theta_{ZZ'}$, and a new coupling of $g_{WWZ'}$ is induced, which will be $\sin\theta_{ZZ'}$ times the SM value, such that $g_{WWZ}^2 + g_{WWZ'}^2= (g_{WWZ}^{SM})^ 2$.  When the CM energy is much larger than $M_{Z'}$ there will be no energy-growing terms in the scattering amplitude.  However, for energies between $M_Z$ and $M_{Z'}$ there will then be effectively $(E/M_W)^4$ growing terms in the scattering amplitude in Eq.~(\ref{amp-gauge}).  If the mass of $Z'$ boson is sufficiently large, the scattering amplitude can enjoy a long period of rise.

  One may argue that the mixing between the SM $Z$ boson and the extra $Z'$ boson arises in the mass matrix of $Z$ and $Z'$.  If the mass of $Z'$ is very heavy, then the mixing angle $\theta_{ZZ'}$ will be extremely small.  In such a case, the growing behavior of $(E/M_W)^4$ term becomes negligible.  The above argument may not be
entirely true.  We will show in the realistic Stueckelberg  $Z'$ model that by choosing a suitable value of $M_{Z'}$ and the mixing angles, the effect of $(E/M_W)^4$ growth can be observed.
We will first demonstrate this heuristically below.
%


In the high energy limit, the amplitudes ${\cal M}_4$ and ${\cal M}^{\rm Higgs}$ for the StSM are the same as their SM counterparts, while
${\cal M}^{\gamma + Z + Z'}_{s,t}$ are given by expressions similar to Eqs.(\ref{ms-sm}) and (\ref{mt-sm}) with $A^{\rm SM}_{x}$ replaced by
the following $A^{\rm StSM}_{x}$
\begin{eqnarray}\label{AxZ'}
A^{\rm StSM}_{x}&=&\frac{s_\theta^2}{x}+\frac{c_\theta^2 c_\psi^2}{x-M_Z^2}+\frac{c_\theta^2 s_\psi^2}{x-M_{Z'}^2}\,.
\end{eqnarray}

Consider first the limit of $x\gg M_{Z'}^2,M_Z^2$ where we have
\begin{equation}\label{Alimit1}
A^{\rm StSM}_{x}
\simeq \frac{1}{x}\bigg[1+\frac{c^2_\theta(M_Z^2  c^2_\psi+M_{Z'}^2 s^2_\psi)}{x}\bigg]\,,\;\;\;(\text{$x \gg M_{Z'}^2,M_Z^2$}) \; .
\end{equation}
In this limit, the sum of the pure gauge diagrams
$i{\cal M}^{\rm gauge} = i \left( {\cal M}_4 + {\cal M}_s^{\gamma + Z + Z'} + {\cal M}_t^{\gamma + Z + Z'} \right)$ is
then
\begin{eqnarray}
\label{amp-gaugeZ-1}
i {\cal M}^{\rm gauge}
& \simeq & - i \frac{g_2^2}{4 M_W^4}\bigg[4M_W^2-3c^2_\theta(M_Z^2  c^2_\psi+M_{Z'}^2 s^2_\psi)\bigg]u + O\left((E/M_W)^0\right) \;
\nonumber\\
& \simeq & - i \frac{g_2^2}{4 M_W^2} u + O\left((E/M_W)^0\right),
\end{eqnarray}
where the mass relation Eq. (\ref{mr2}) has been used. This limit is the same as the SM and thus when combined with the Higgs contribution
the total amplitude is well behaved at high energy.

Next consider the intermediate range of $M_{Z}^2 <x \ll M_{Z'}^2$ where we have
\begin{eqnarray}\label{Alimit2}
A^{\rm StSM}_{x}
&\simeq& \frac{1}{x}\bigg[1-c^2_\theta s^2_\psi \left(1-\frac{M_Z^2}{x}\right)^{-1}+\frac{M_Z^2 c^2_\theta}{x-M_Z^2}\bigg] \; ,
\;\;\;(M_{Z}^2 < x \ll M_{Z'}^2) \; .
\end{eqnarray}
In this limit the sum of all gauge diagrams is given by
\begin{equation}
\label{amp-gaugeZ-2}
i {\cal M}^{\rm gauge} \simeq   i \frac{g_2^2}{4 M_W^4}c^2_\theta s^2_\psi (s^2 + 4 st + t^2)+ O\left((E/M_W)^{2}\right) \; .
\end{equation}
This sum of pure gauge amplitudes is of order $(E/M_W)^4$ and therefore cannot be cancelled by the Higgs contribution which is of order
$(E/M_W)^2$. If the factor of the mixing angles $c^2_\theta s^2_\psi$ is not too small in the intermediate range,
there should be discernible effects.
We note that there is a similar $(E/M_W)^4$ growth in the partial decay
width of $Z'\to W^+W^-$ \cite{Z'toWW} with $E=M_{Z'}$.
This growth is most relevant for $W^+ W^- \longrightarrow W^+ W^-$ scattering when
$s \sim M_{Z'}^2$, at which the scattering amplitude factorizes and the cross section is proportional to
$\left( \Gamma^{Z'}_{W^+W^-}\right )^2 / \left( M_{Z'}^2 \Gamma^2_{Z'} \right)$.


\subsection{Exact Scattering Amplitude for $W^+_L W^-_L  \to W^+_L W^-_L$}

The exact scattering amplitude for the process $W_L^-(p_1, \mu) \, W_L^+(p_2, \lambda) \to W_L^-(p_3, \nu) \, W_L^+(p_4, \rho)$ in the StSM reviewed in the previous section can be easily derived.  It can be expressed as
\begin{eqnarray}
{\cal M}&=& g_2^2 \Biggr \{  \left[2 g_{\mu\rho} g_{\nu\lambda}
  - g_{\mu\nu} g_{\rho\lambda}   - g_{\mu\lambda} g_{\nu\rho} \right]  \nonumber \\
&+& \left( \frac{s_\theta^2 }{s} + \frac{ c_\theta^2 c_\psi^2}{s-M_Z^2}
     + \frac{c_\theta^2 s_\psi^2}{ s - M_{Z'}^2 + i M_{Z'} \Gamma_{Z'} } \right )\;
  \biggr[ (-2p_1 -p_2)_\lambda g_{\alpha\mu} + (p_1 - p_2)_\alpha g_{\mu\lambda}
   \nonumber \\
&&  + (p_1 +2 p_2)_\mu g_{\lambda\alpha} \biggr] \biggr [
  (p_3 + 2 p_4)_\nu g_\rho^\alpha + (p_3 - p_4)^\alpha g_{\nu\rho}
  +( -2p_3 - p_4)_\rho g^\alpha_\nu \biggr] \nonumber \\
&+& \left( \frac{s_\theta^2 }{t} + \frac{ c_\theta^2 c_\psi^2}{t-M_Z^2}
     + \frac{c_\theta^2 s_\psi^2}{ t - M_{Z'}^2  } \right )\;
  \biggr[ (-2p_1 +p_3 )_\lambda g_{\alpha\mu} +
(p_1 + p_3)_\alpha g_{\mu\nu}
   \nonumber \\
&&  + (p_1 -2 p_3)_\mu g_{\nu\alpha} \biggr] \biggr [
  (-p_2 + 2 p_4)_\lambda g_\rho^\alpha + (-p_2 - p_4)^\alpha g_{\lambda\rho}
  +( 2p_2 - p_4)_\rho g^\alpha_\lambda \biggr] \nonumber \\
&-&  M_W^2 \left(  \frac{1}{s - M_h^2} g_{\mu\lambda} g_{\nu\rho}
                      + \frac{1}{t - M_h^2} g_{\mu\nu} g_{\lambda\rho} \right )
 \Biggr \}\, \epsilon_L^\mu(p_1)  \, \epsilon_L^\lambda(p_2) \, \epsilon_L^{*\nu}(p_3)
\, \epsilon_L^{*\rho}(p_4)  ~,
\end{eqnarray}
where the longitudinal polarization vector $\epsilon_L^\mu (p_i) = (\vert {\bf p}_i \vert , E_i {\hat {\bf p}_i} ) / M_W$ for $p_i^\mu = ( E_i, \vert {\bf p}_i \vert {\hat {\bf p}_i} )$.
Formulas for the other $WW$ scattering processes can be worked out straightforwardly as well.  Since their expressions are not illuminating, we will not present them here.  With these scattering amplitudes in hand, we can then fold them with the parton distribution functions as well as the $WW$ luminosity to obtain the scattering cross sections for $pp \to WW jj + X$ at the LHC.  We note that the enhancement due to the incomplete cancellation is not at all obvious because of the reduction in the parton probabilities at high $x$.  Detailed numerical studies are required, which we will turn to in the next section.

To study unitarity constraints, we need to project out the partial wave coefficients $a^I_J$ for different channels with total angular momentum $J$ and isospin $I$ from the above scattering amplitudes.  The partial wave coefficients for the dominant $S$- and
$P$-wave scatterings are given by
\begin{eqnarray}
\label{a00}
a^0_0 &=& \frac{1}{64\pi} \int^1_{-1} d \cos\theta \, \left[
3\, {\cal M}\left ( W^+_L W^-_L \to Z_L Z_L \right )
+ {\cal M} \left(W^+_L W^+_L \to W^+_L W^+_L \right ) \right ]
\;, \\
\label{a11}
a^1_1 &  =  & \frac{1}{64\pi} \int^1_{-1} d \cos\theta \, \left\{
2 \left[ {\cal M}\left(W^+_L W_L^- \to W^+_L W_L^- \right )
 -  \, {\cal M}\left ( W^+_L W^-_L \to Z_L Z_L \right ) \right] \right. \nonumber \\
&&  \left. - \, {\cal M} \left( W^+_L W^+_L \to W^+_L W^+_L \right ) \right\} \cos \theta  \; , \\
\label{a20}
a^2_0 & =& \frac{1}{64\pi} \int^1_{-1} d \cos\theta \, {\cal M}\left (
W^+_L W^+_L \to W^+_L W^+_L \right ) \; .
\end{eqnarray}
Unitarity requires $|\Re e\; a^I_J| \le 1/2$, in particular for $ a^0_0$, $ a^1_1$ and $ a^2_0$ that we are interested.


\section{Numerical Results}

\begin{figure}[th!]
\centering
\includegraphics[width=3in]{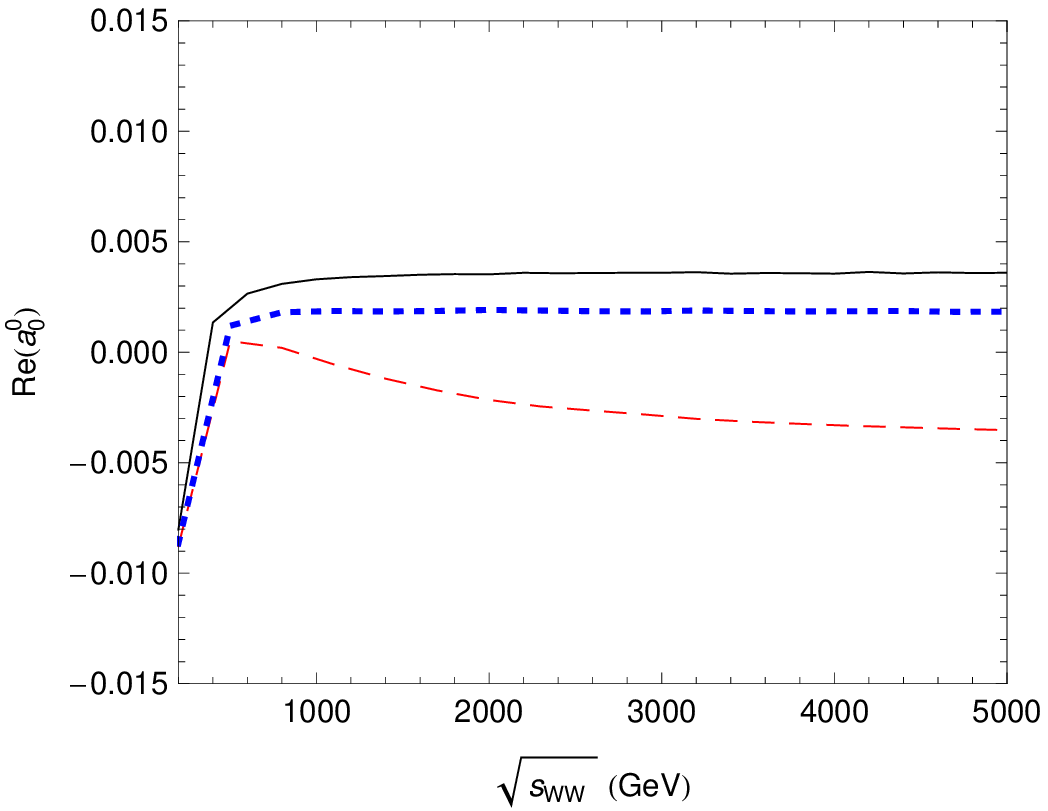}
\includegraphics[width=3in]{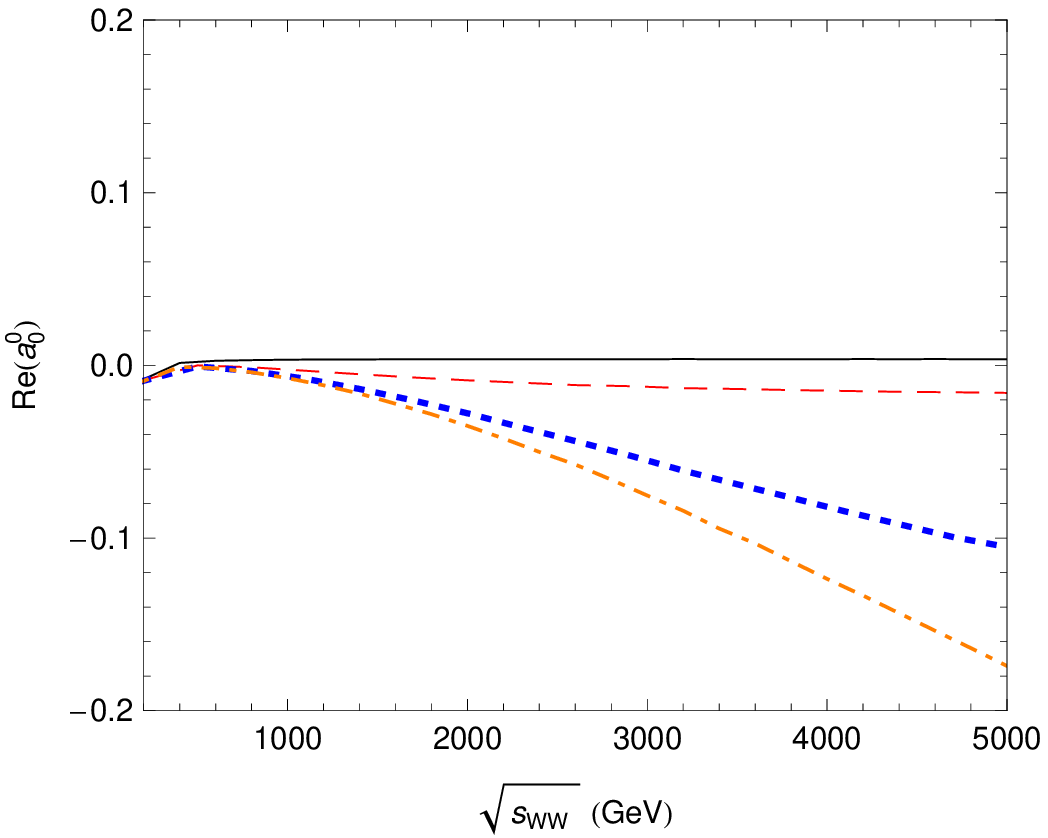}
\includegraphics[width=3in]{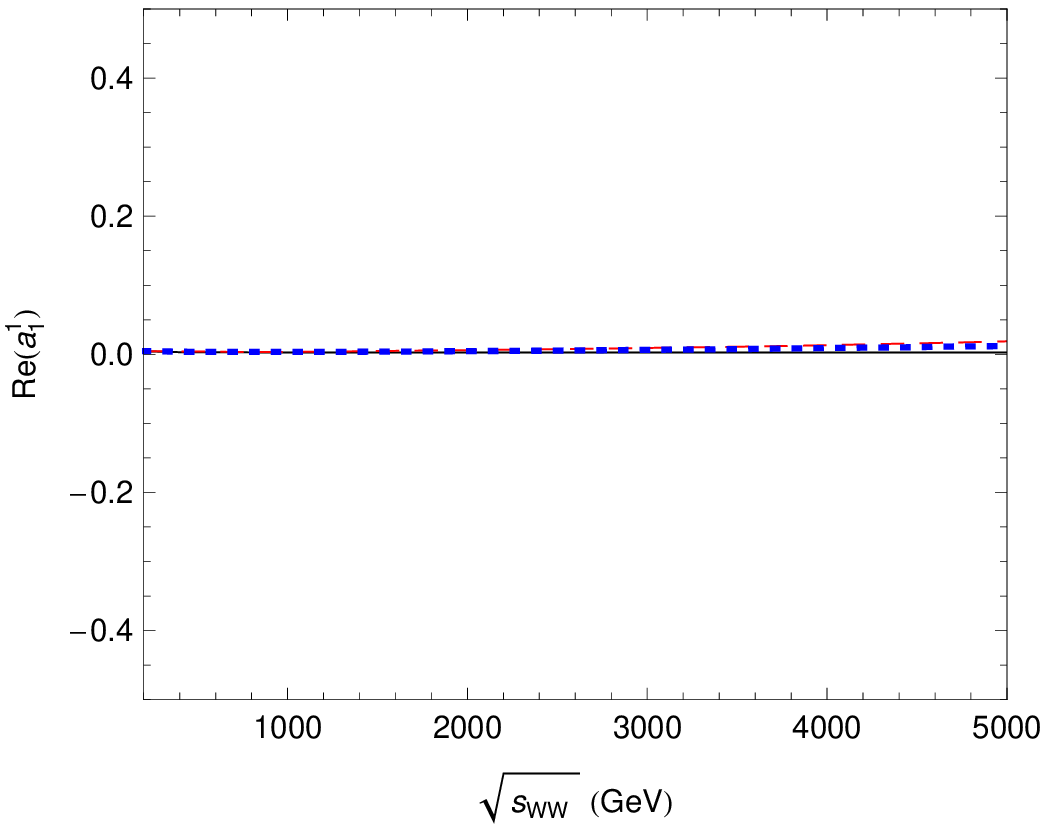}
\includegraphics[width=3in]{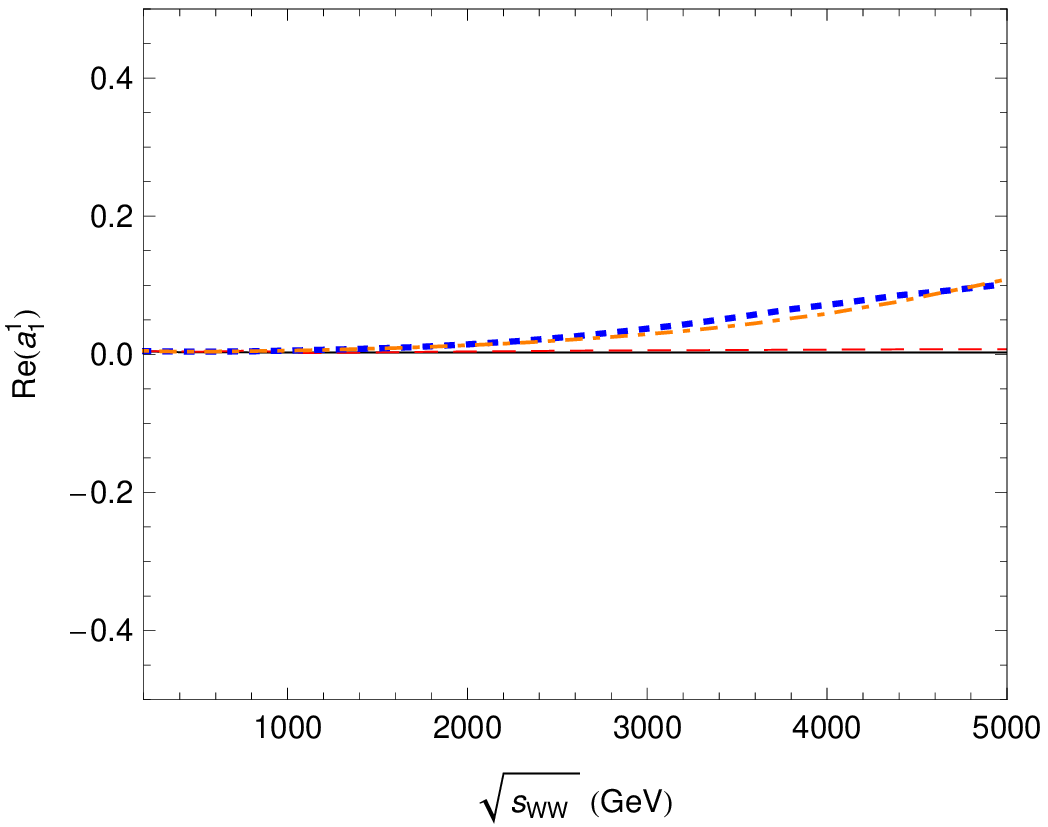}
\includegraphics[width=3in]{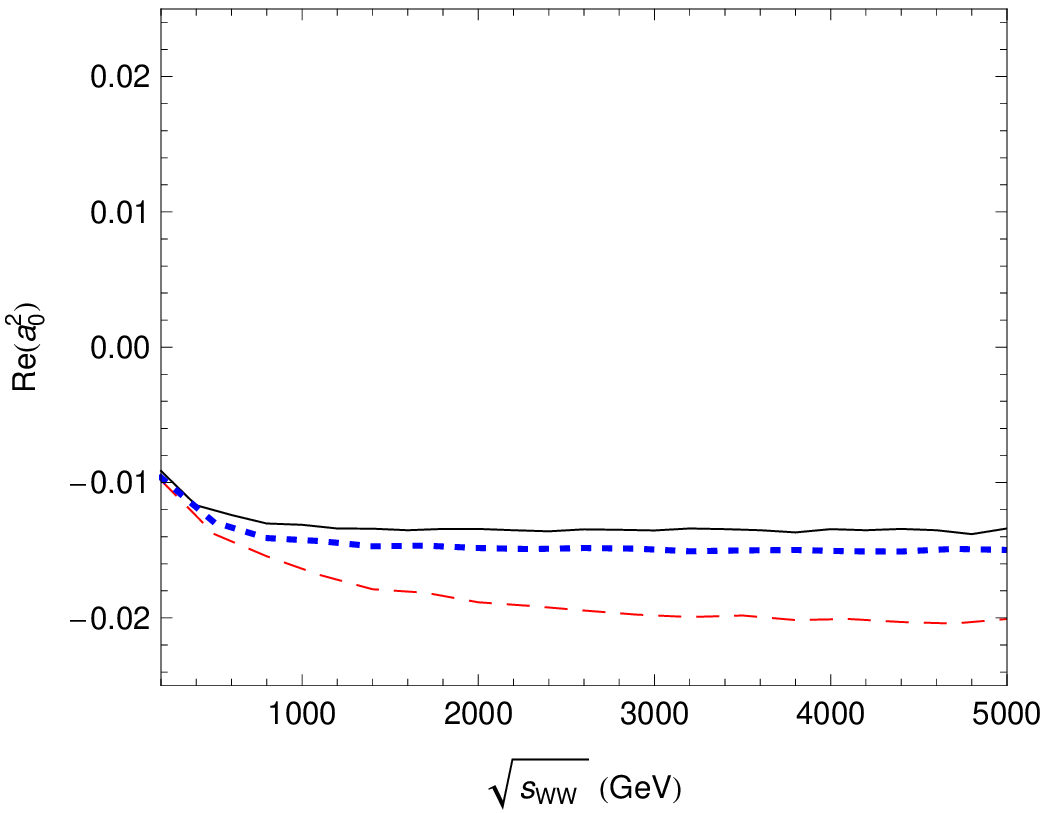}
\includegraphics[width=3in]{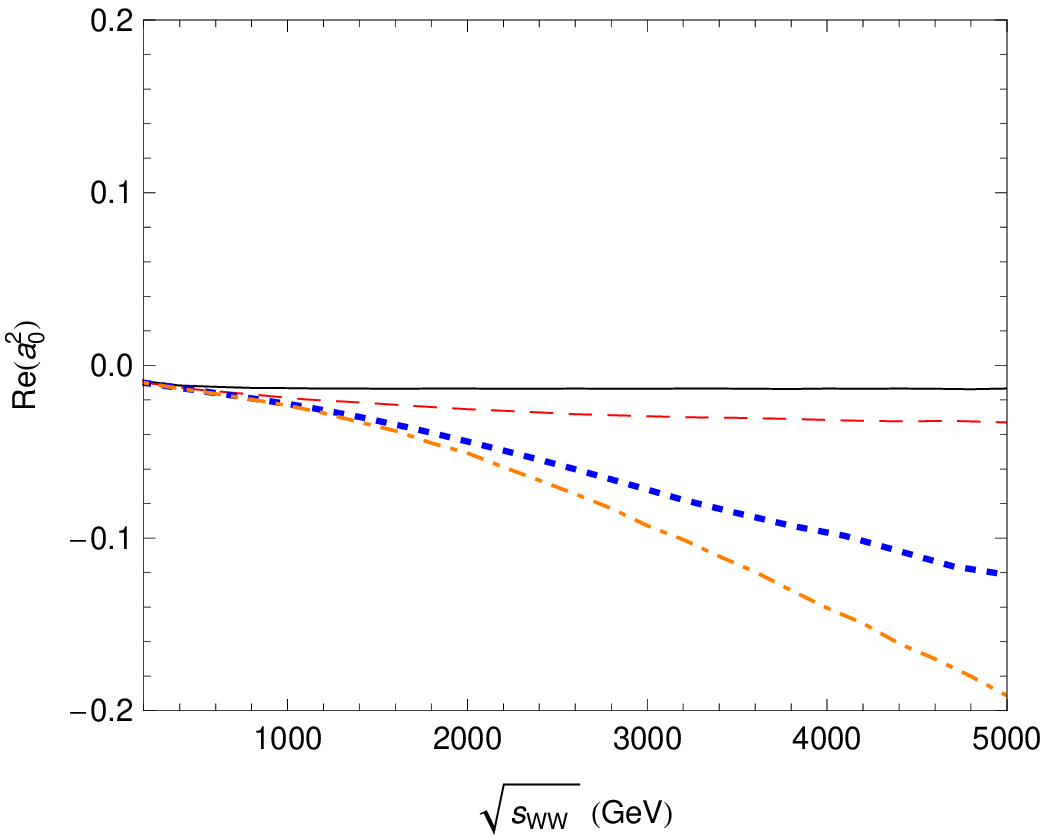}
\caption{The partial wave coefficients $\Re e\; a^0_{0}$, $\Re e\; a^1_{1}$ and $\Re e\; a^2_{0}$ versus the scattering energy $\sqrt{s_{WW}}$.  Left: The solid line is for the SM, while the dash (dotted) line denotes $M_{Z'}$ = 600 (300) GeV with $\epsilon$ = 1.73 (0.77) for the StSM.  Right: The solid line is for the SM, while the dash, dotted and dotted-dash lines denote $M_{Z'}$ = 1, 3, 5 TeV with $\epsilon$ = 3.05, 10.13, 17.59, respectively, for the StSM.}
\label{partialwave}
\end{figure}

In Fig.~\ref{partialwave}, we plot the partial wave coefficients $a_J^I$ as a function of CM energy $\sqrt{s_{WW}}$ for $a^0_0$, $ a^1_1$ and $ a^2_0$
from top to bottom, respectively.  The solid lines in these plots are the SM results.
In the left panel, the dash line has $M_{Z'}$ = 600 GeV and $\epsilon$ = 1.73, while the dotted line has $M_{Z'}$ = 300 GeV and $\epsilon$ = 0.77.
It is clear that the results for the StSM have no difference from the SM ones for these choices of parameters.
In the right panel, the dash, dotted and dotted-dash lines have $M_{Z'} = 1,3$ and 5 TeV with $\epsilon$ taken to be 3.05, 10.13 and 17.59 respectively.
These deviations from the SM results are due to the incomplete cancellation of the bad high energy terms that we have alluded to erstwhile.  However, the growth in the partial wave coefficients do not post any threats to the unitarity limit of 0.5 even for such a large $\epsilon$ scenario.
We note that this large angle scenario is motivated by the contour plot in Fig. \ref{zmasscontour} which suggests that
large $\epsilon$ values are possible provided that the $Z'$ mass is also large so as not to upset the experimental value of
the $Z$ mass. It is therefore quite interesting to repeat
the global fit analysis done in \cite{Feldman-Liu-Nath-1} where a small $\epsilon$ was assumed. We would like to relegate
such analysis to a future work.

\begin{figure}[th!]
\includegraphics[width=3in]{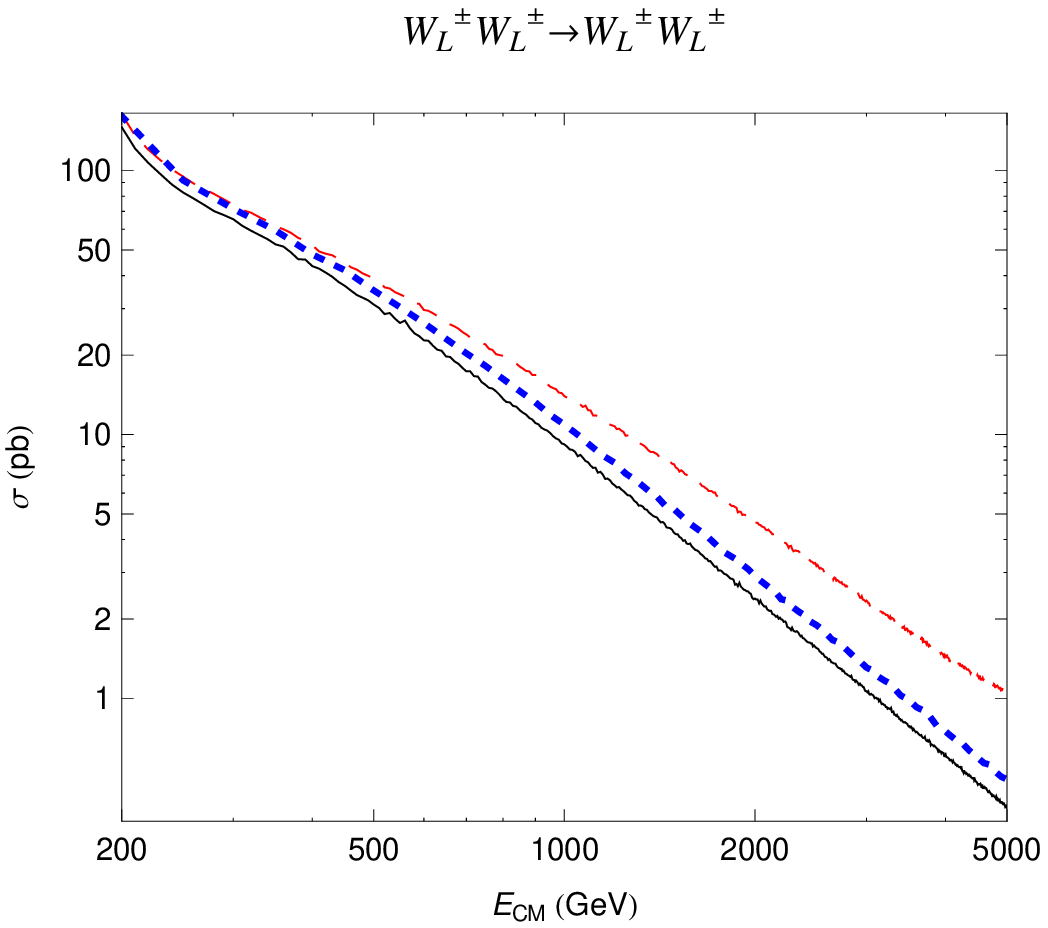}
\includegraphics[width=3in]{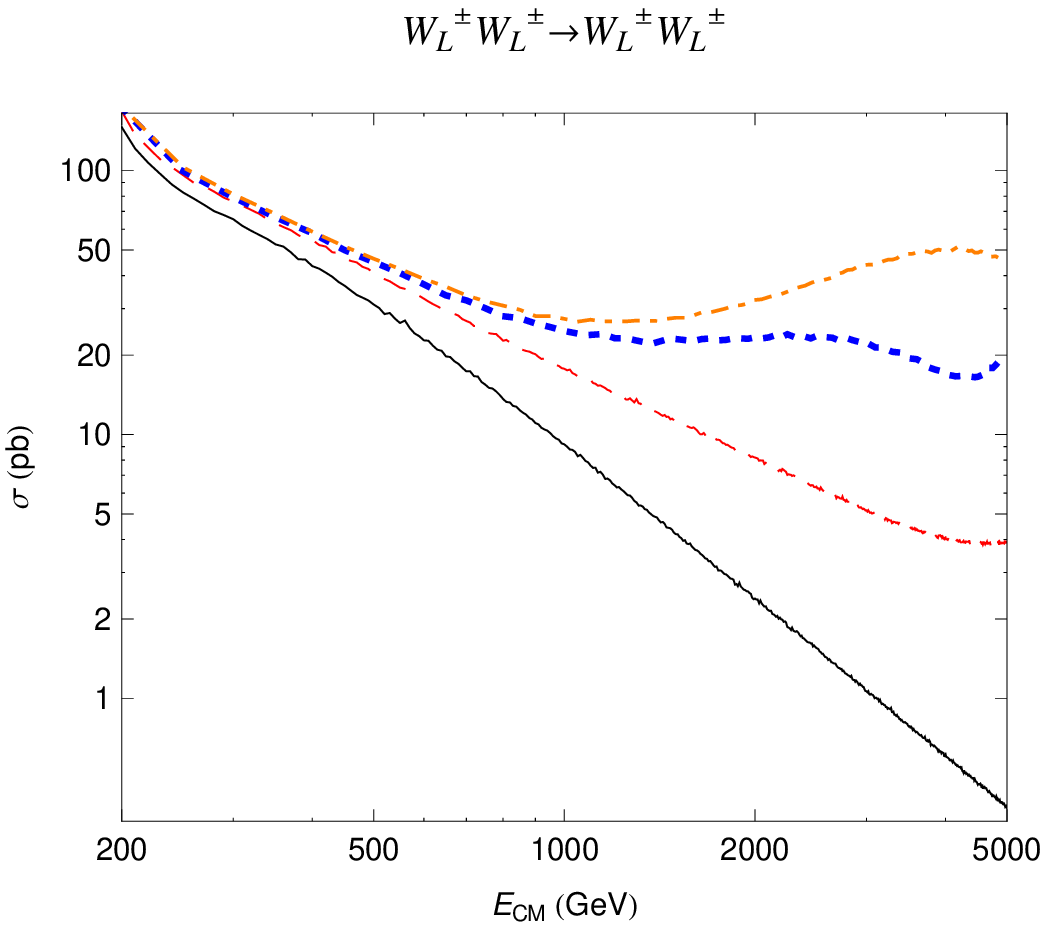}
\includegraphics[width=3in]{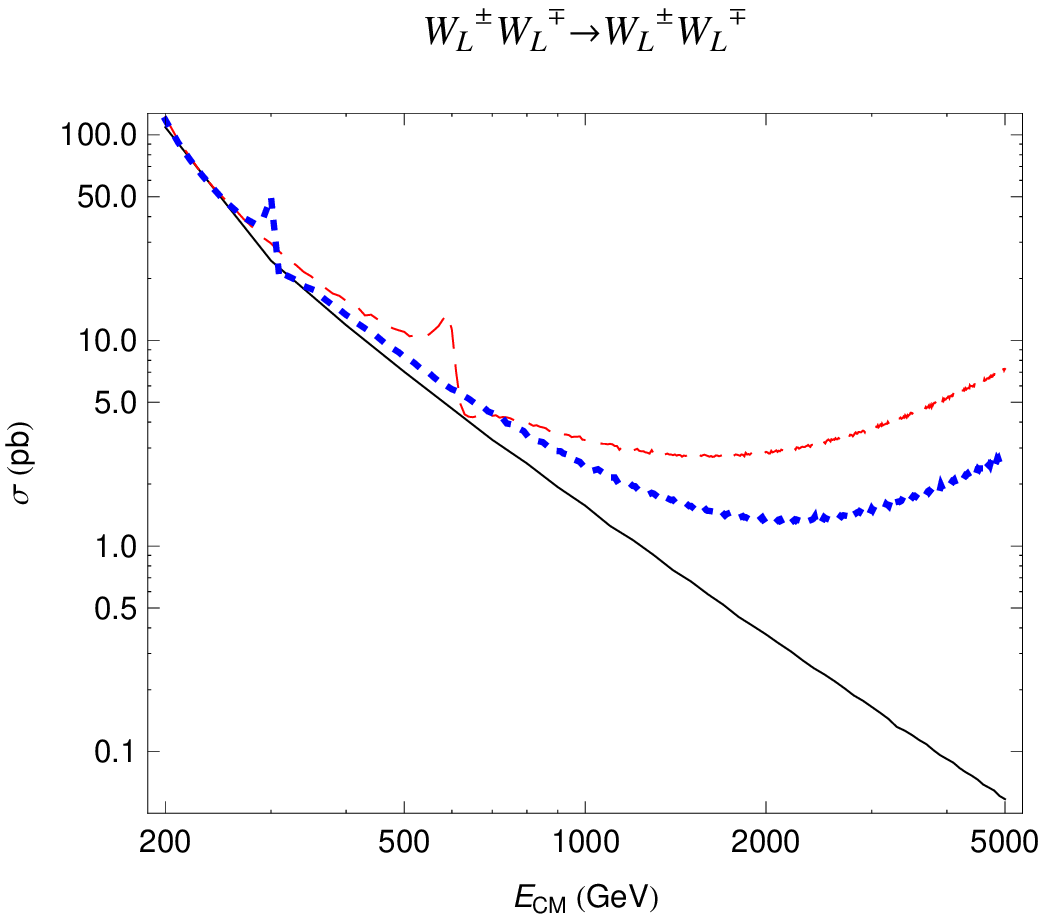}
\includegraphics[width=3in]{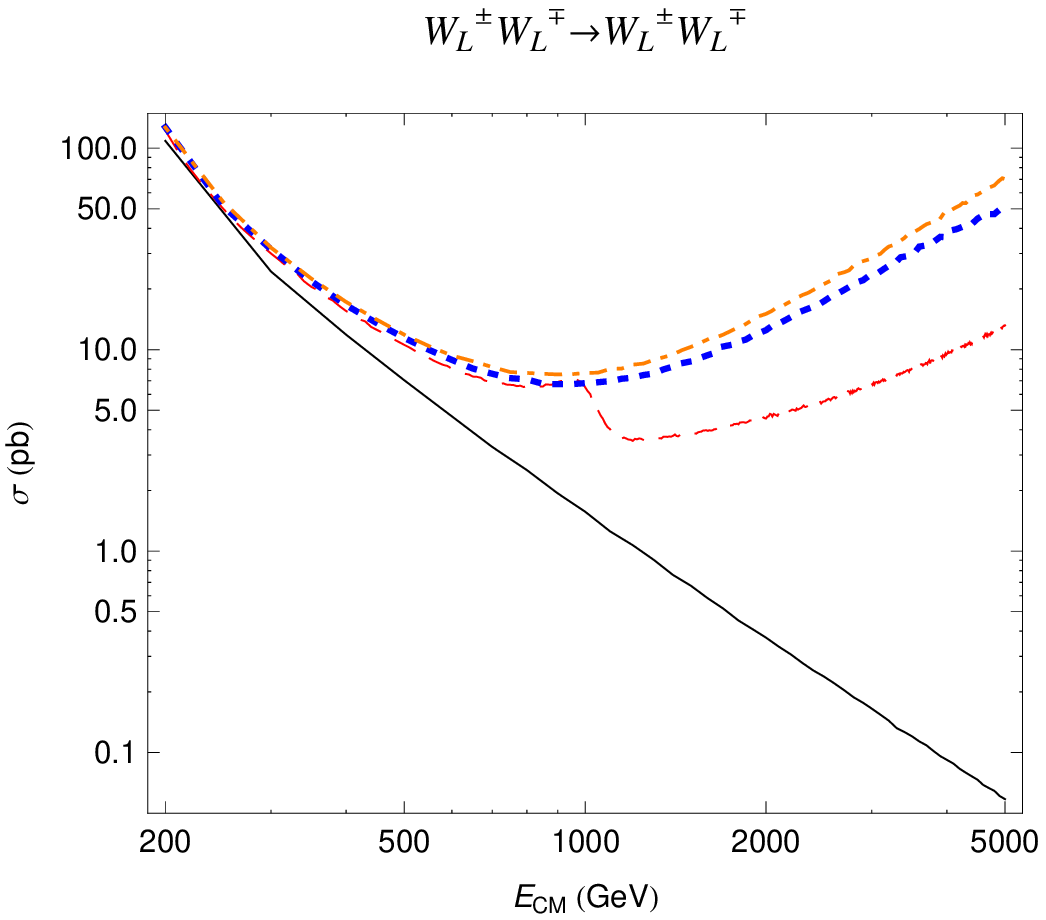}
\caption{Total cross sections for the two processes of $W^\pm_L W^\pm_L \to W^\pm_L W^\pm_L$ and $W^\pm_L W^\mp_L \to W^\pm_L W^\mp_L$ versus the center-of-mass energy.  Left: The solid line is for the SM, while the dash (dotted) line denotes $M_{Z'}$ = 600 (300) GeV with $\epsilon$ = 1.73 (0.77) for the StSM.  Right: The solid line is for the SM, while the dash, dotted and dotted-dash lines denote $M_{Z'}$ = 1, 3, 5 TeV with $\epsilon$ = 3.05, 10.13, 17.59, respectively, for the StSM.}
\label{WWscattering}
\end{figure}

\begin{figure}[th!]
\includegraphics[width=3in]{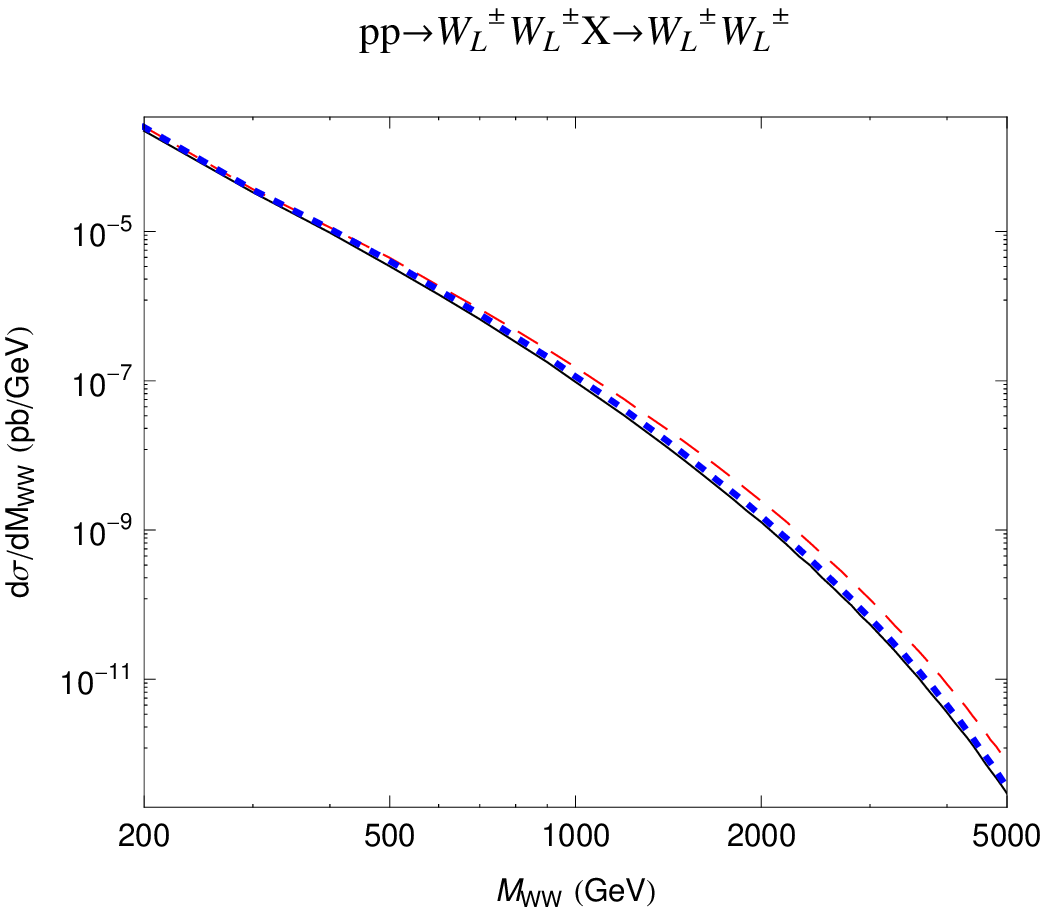}
\includegraphics[width=3in]{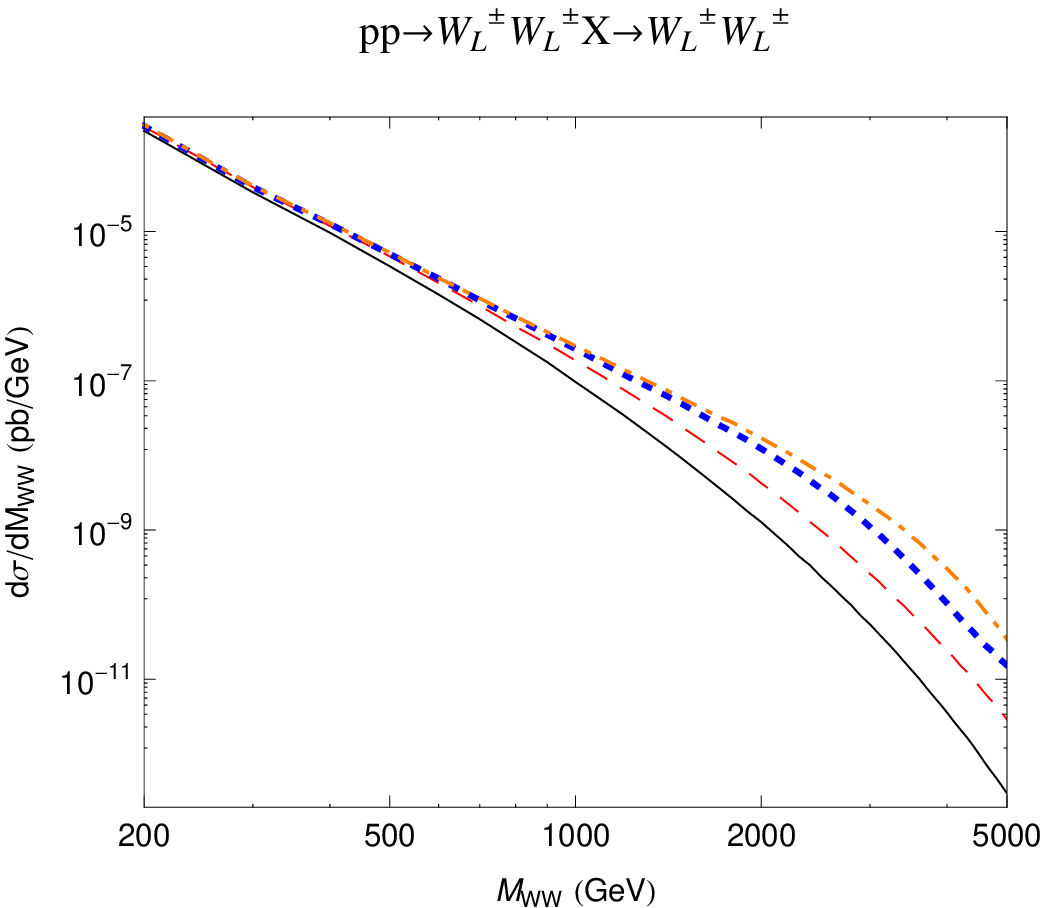}
\includegraphics[width=3in]{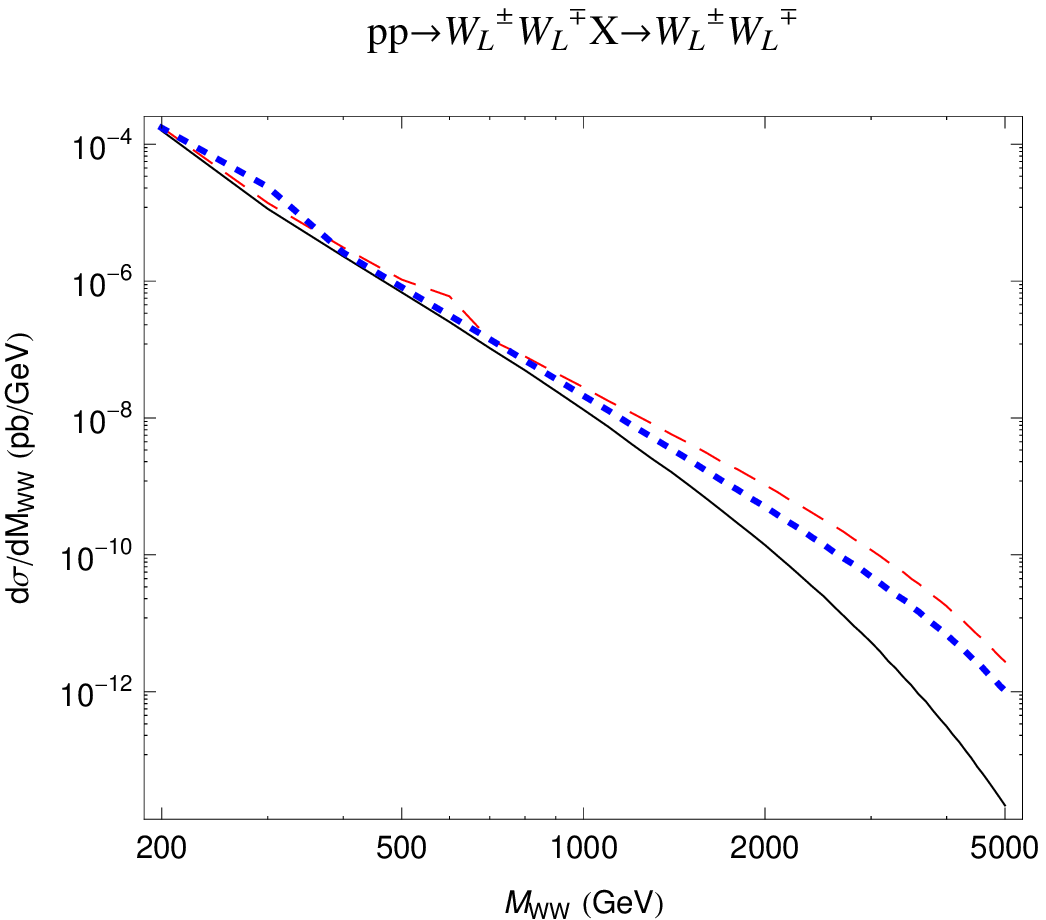}
\includegraphics[width=3in]{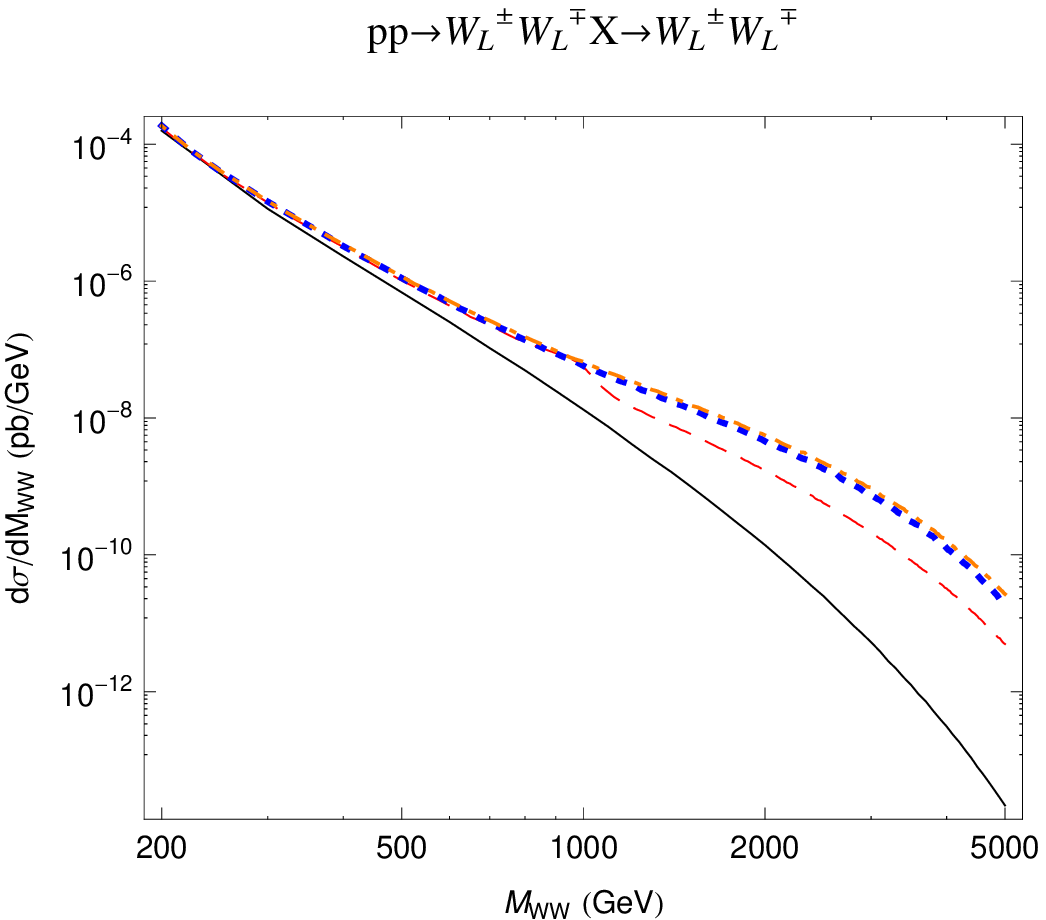}
\caption{Invariant mass distributions for the two processes of $W^\pm_L W^\pm_L \to W^\pm_L W^\pm_L$ and $W^\pm_L W^\mp_L \to W^\pm_L W^\mp_L$ at the LHC.  Left: The solid line is for the SM, while the dash (dotted) line denotes $M_{Z'}$ = 600 (300) GeV with $\epsilon$ = 1.73 (0.77) for the StSM.  Right: The solid line is for the SM, while the dash, dotted and dotted-dash lines denote $M_{Z'}$ = 1, 3, 5 TeV with $\epsilon$ = 3.05, 10.13, 17.59, respectively, for the StSM.}
\label{ppscattering}
\end{figure}

In Fig.~\ref{WWscattering}, we plot the total cross sections for the two most relevant processes $W^\pm_L W^\pm_L \to W^\pm_L W^\pm_L$ (top panel) and $W^\pm_L W^\mp_L \to W^\pm_L W^\mp_L$ (bottom panel) as a function of the CM energy.  The legends for the left and right panels of these plots are the same as those in Fig.~\ref{partialwave}.  The corresponding growth of the total cross section is evident from the plots in the right panel for the large $\epsilon$ scenario.  In Fig.~\ref{ppscattering}, we plot the differential cross sections for $pp \to W^\pm_L W^\pm_L+X$ (top panel) and $pp \to W^\pm_L W^\mp_L+X$ (bottom panel) as a function of the invariant mass of the $WW$ boson pair by folding with the parton luminosities at the LHC.  Again, the legends for the left and right panels of these plots are the same as those in Fig.~\ref{partialwave}.  Due to the suppression from the parton luminosities at large $x$, the enhancement seen from the partial wave coefficients of Fig.~\ref{partialwave} is less obvious at the hadronic level.

\begin{table*}[th!]
  \caption{Event rates for longitudinal weak gauge bosons scattering at the LHC with an assumed yearly luminosity of 100 fb$^{-1}$.  Branching ratios for the leptonic final states are summed over for $\ell = e$ and $\mu$. We set $M_h = 120$ GeV, $M^{\text{min}}_{WW}=200$ GeV, $\vert \cos \theta^*_{WW} \vert \leq 0.8$
and $\alpha_{em}(M_{Z'})$.}
\label{eventrates}
\begin{ruledtabular}
\begin{tabular}{l|cccccc}
$\quad\quad\quad\quad$
Subprocess
& $\text{SM}$
& $^{M_{Z'}\, = \, 300\,\text{GeV}}_{\;\;\;\;\;\;\epsilon \, = \, 0.77}$
& $^{M_{Z'}\, = \, 600\,\text{GeV}}_{\;\;\;\;\;\;\epsilon \, = \, 1.73}$
& $^{M_{Z'}\,=\,1\,\text{TeV}}_{\;\;\;\;\;\;\epsilon \,=\,3.05}$
& $^{M_{Z'}\,=\,3\,\text{TeV}}_{\;\;\;\;\;\;\epsilon \,=\,10.13}$
& $^{M_{Z'}\,=\,5\,\text{TeV}}_{\;\;\;\;\;\;\epsilon \,=\,17.59}$\\
\hline
$W^{\pm}_L W^{\pm}_L \to  W^{\pm}_L W^{\pm}_L \to \ell^\pm \nu \ell^\pm  \nu$&50.97    &56.62   &58.77    &60.35   &63.04  &64.27 \\\hline
$W^{\pm}_L W^{\mp}_L \to  W^{\pm}_L W^{\mp}_L \to \ell^\pm \nu \ell^\mp \nu$ &25.67    &27.54   &28.17    &28.55   &29.54  &30.29 \\\hline
$W^{\pm}_L Z_L \to  W^{\pm}_L Z_L \to \ell^\pm \nu \ell^+\ell^- $            &5.50     &5.30    &5.30     &5.30    &5.30   &5.30  \\\hline
$W^{\pm}_L W^{\mp}_L \to  Z_L Z_L \to \ell^+\ell^- \ell^+ \ell^-$            &0.42     &0.41    &0.42     &0.42    &0.42   &0.42  \\
$W^{\pm}_L W^{\mp}_L\to  Z_L Z_L \to \ell^+\ell^-  \nu \bar \nu$             &2.50     &2.46    &2.50     &2.50    &2.50   &2.51  \\\hline
$Z_L Z_L \to  W_L^\pm W_L^\mp \to \ell^\pm \nu \ell^\mp \nu $                &3.11     &2.98    &3.00     &3.01    &3.04   &3.04 \\
$Z_L Z_L \to  Z_L Z_L \to \ell^+ \ell^- \ell^+ \ell^-$                       &0.13     &0.12    &0.12     &0.12    &0.11   &0.11  \\
$Z_L Z_L \to  Z_L Z_L \to \ell^+ \ell^- \nu \bar \nu$                        &0.79     &0.71    &0.70     &0.69    &0.67   &0.66 \\
\end{tabular}
\end{ruledtabular}
\end{table*}
\noindent

In Table~\ref{eventrates}, we present the event rates for the various longitudinal weak gauge boson scattering processes at the LHC for the StSM with the same parameter choices as in the previous figures.
Here, as well as in previous figures, we have imposed the kinematic cut $\vert \cos \theta^*_{WW} \vert \leq 0.8$
at the parton rest frame. The SM results are also shown for comparison.
As one can see, the rise for the $W^\pm_L W^\pm_L \to W^\pm_L W^\pm_L$ and $W^\pm_L W^\mp_L \to W^\pm_L W^\mp_L$ channels are discernible for the large $\epsilon$ scenario.

\section{Conclusions}

In models with an additional heavy neutral gauge boson, modifications of the trilinear and quartic pure gauge couplings and the gauge-Higgs couplings are possible through the mixings among the SM $Z$, the extra $Z'$ and possibly the photon as well.  In this work, using the simple Stueckelberg extension of the SM as an example, we demonstrate that these modifications can lead to the enhancement of the partial wave coefficients of the longitudinal weak gauge boson scatterings as compared with the SM.  However, this phenomenon occurs only with the large mixing angle scenario.
While we are fully aware of our choices of the parameter values for the large $\epsilon$ scenario might not be realistic, they are sufficient to demonstrate $W_LW_L$ scatterings as a sensitive probe to
both the pure gauge structure as well as the electroweak symmetry breaking mechanism.
Thus, it should be interesting to study if this scenario is consistent with existing experimental constraints from LEP and Tevatron by performing the global fits for the Stueckelberg extension of the SM with or without the kinetic mixing term.

In many extensions of the SM, other types of extra $U(1)$ gauge groups are possible.  These include sequential $Z'$, superstring $Z'$ \cite{Ringwald,Guzzi} and various types of $Z'$ based on $E_6$ unification \cite{erler}.  Data from electroweak precision tests, LEP II and CDF/D0 had put stringent constraints on both the mixing angle as well as the $Z'$ mass for these models \cite{erler}.  Thus they are similar to the StSM with small mixing angles that we have also studied in this work.  Hence there should be no difference from the SM for the longitudinal $WW$ scatterings in these models.

Before closing, we note that an extra $W'$ mixing with the SM $W$ may also lead to modifications of the trilinear and quartic couplings in the pure gauge sector as well as the $hWW$ couplings. They may give rise to enhancement in the other channels like $W_L^\pm W_L^\mp \leftrightarrow Z_L Z_L$ and $Z_L W_L^\pm \to Z_L W_L^\pm$ which are shown in our analysis to have no difference from the SM results for the StSM even with large mixing angles.
Thus in general one should bear in mind that scatterings of longitudinal weak gauge bosons are not only sensitive to the underlying electroweak symmetry breaking mechanism, but also to the pure gauge sector structure.


\section*{Acknowledgement}
The work was supported in parts by the National Science Council of Taiwan under Grant Nos. 96-2628-M-007-002-MY3, 97-2112-M-008-002-MY3 and 98-2112-M-001-014-MY3, the NCTS, the Boost Program of NTHU and the WCU program through the NRF funded by the MEST (R31-2008-000-10057-0).


\end{document}